\documentclass[conference]{IEEEtran}
\ifCLASSINFOpdf
\else
\fi
\usepackage{amsmath,dsfont,bbm,epsfig,amssymb,amsfonts,amstext,verbatim,amsopn,cite,subfigure,multirow,multicol,lipsum,xfrac}
\usepackage{amsthm}
\usepackage{mathtools,amsthm}
\usepackage{perpage}
\usepackage{balance}
\usepackage{url}
\usepackage{amsfonts}
\usepackage{epsfig}
\usepackage[font={small}]{caption}
\usepackage{psfrag}	
\usepackage{etoolbox}
\usepackage{algorithmicx}
\usepackage[Algorithm,ruled]{algorithm}
\usepackage{algpseudocode}
\usepackage{pifont}
\usepackage[utf8]{inputenc}
\usepackage[T1]{fontenc}  
\usepackage[nolist]{acronym}
\MakePerPage{footnote}
\usepackage{paralist}
\usepackage{enumitem}
\usepackage{bbm}
\usepackage[process=auto]{pstool}
\usepackage{tikz}
\usetikzlibrary{shapes,arrows}
\usepackage{color}
\definecolor{myblue}{rgb}{0.07, 0.41, 0.7}
\definecolor{myred}{rgb}{0.85, 0.33, 0.1}
\newcommand{\setX}{\mathbbmss{X}}

\newcommand{\setP}{\mathbbmss{P}}
\newcommand{\setR}{\mathbbmss{R}}

\newcommand{\setW}{\mathbbmss{W}}

\newcommand{\setZ}{\mathbbmss{Z}}

\newcommand{\e}{\mathrm{e}}

\newcommand{\rmp}{\mathrm{p}}

\newcommand{\rmq}{\mathrm{q}}

\newcommand{\rmR}{\mathrm{R}}
\newcommand{\rmg}{\mathrm{g}}
\newcommand{\rmG}{\mathrm{G}}

\newcommand{\sfQ}{\mathsf{Q}}

\newcommand{\sfD}{\mathsf{D}}

\newcommand{\sfd}{\mathsf{d}}

\newcommand{\sfp}{\mathsf{p}}

\newcommand{\maf}{\mathcal{F}}

\newcommand{\mae}{\mathcal{E}}
\newcommand{\maz}{\mathcal{Z}}

\newcommand{\man}{\mathcal{N}}

\newcommand{\bxx}{\mathbf{x}}
\newcommand{\byy}{\mathbf{y}}

\newcommand{\bvv}{\mathbf{v}}

\newcommand{\bx}{{\boldsymbol{x}}}
\newcommand{\hx}{{\hat{x}}}

\newcommand{\vv}{\mathrm{v}}
\newcommand{\hxx}{\hat{\mathrm{x}}}
\newcommand{\xx}{\mathrm{x}}
\newcommand{\yy}{\mathrm{y}}
\newcommand{\zz}{\mathrm{z}}

\newcommand{\bhx}{{\boldsymbol{\hat{x}}}}
\newcommand{\set}[1]{\left\lbrace#1\right\rbrace}

\newcommand{\bz}{{\boldsymbol{z}}}

\newcommand{\bv}{{\boldsymbol{v}}}

\newcommand{\by}{{\boldsymbol{y}}}

\newcommand{\trp}{\mathsf{T}}

\newcommand{\mA}{\mathbf{A}}
\newcommand{\mR}{\mathbf{R}}

\newcommand{\mI}{\mathbf{I}}
\newcommand{\mone}{\mathbf{1}}
\newcommand{\mJ}{\mathbf{J}}

\newcommand{\mQ}{\mathbf{Q}}

\newcommand{\mU}{\mathbf{U}}
\newcommand{\mD}{\mathbf{D}}
\newcommand{\mM}{\mathbf{M}}

\newcommand{\mT}{\mathbf{T}}
\newcommand{\mH}{\mathbf{H}}

\newcommand{\E}[1]{\mathbb{E} \left\lbrace #1 \right\rbrace }

\newcommand{\mse}{\mathsf{mse}}

\def\argmin{\mathop{\rm argmin}}

\newcommand{\norm}[1]{\lVert #1 \rVert}

\newcommand{\img}[1]{\mathsf{Im}\left\lbrace #1 \right\rbrace}

\newcommand{\abs}[1]{\lvert #1 \rvert}

\newtheoremstyle{mystyle}
  {}
  {}
  {}
  {}
  {\bfseries}
  {:}
  { }
  {}

\theoremstyle{mystyle}

\newcounter{bar}

\begin{document}
\title{Theoretical Bounds on MAP Estimation in Distributed Sensing Networks\vspace*{-3mm}}
\author{
\IEEEauthorblockN{
Ali Bereyhi\IEEEauthorrefmark{1},
Saeid Haghighatshoar\IEEEauthorrefmark{2},
Ralf R. M\"uller\IEEEauthorrefmark{1},
}
\IEEEauthorblockA{
\IEEEauthorrefmark{1}Institute for Digital Communications (IDC), Friedrich-Alexander Universit\"at Erlangen-N\"urnberg\\
\IEEEauthorrefmark{2}Communications and Information Theory Group (CommIT), Technische Universit\"at Berlin\\
ali.bereyhi@fau.de, 
saeid.haghighatshoar@tu-berlin.de, 
ralf.r.mueller@fau.de\vspace*{-4mm} 
\thanks{This work was supported by the German Research Foundation, Deutsche Forschungsgemeinschaft (DFG), under Grant No. MU 3735/2-1.}
}
}

\IEEEoverridecommandlockouts

\maketitle
\tikzstyle{block} = [draw, rounded corners, rectangle, minimum height=2.5em, minimum width=5em]
\tikzstyle{margin} = [draw, dotted, rectangle, minimum height=2.1em, minimum width=2em]
\tikzstyle{sum} = [draw, circle, node distance=1cm, inner sep=0pt]
\tikzstyle{input} = [coordinate]
\tikzstyle{output} = [coordinate]
\tikzstyle{pinstyle} = [pin edge={to-,thick,black}]
\begin{abstract}
The typical approach for recovery of spatially~correlated signals is regularized least squares with a~coupled~regula-rization term. In the Bayesian framework, this algorithm~is~seen as a maximum-a-posterior estimator whose postulated prior is proportional to the regularization term. In this paper, we study distributed sensing networks in which a set of spatially correlated signals are measured individually at separate terminals,~but~recovered jointly via a generic maximum-a-posterior estimator.~Using the replica method, it is shown that the setting exhibits the decoupling property. For the case with jointly sparse signals, we invoke Bayesian inference and propose~the~``multi-dimensional soft thresholding'' algorithm which is posed as a~linear program-ming. Our investigations depict that the~proposed~algorithm~out-performs the conventional $\ell_{2,1}$-norm regularized least squares scheme while enjoying a feasible computational complexity.\vspace*{1mm}
\end{abstract}

\IEEEpeerreviewmaketitle

\section{Introduction}
In a \ac{dsn}, the receiving~ter- minal deals with the problem of signal recovery from~a~set~of individually measured observations which are generally~underdetermined and noisy. Common examples of such networks arise in \ac{dcs} \cite{baron2005information,duarte2005joint} and the~\ac{mmv}~problem~\cite{cotter2005sparse}. The typical approach for signal recovery in \ac{dsn}s is~\ac{rls} regression whose performance, as well as complexity, depends on the regularization term. In a Bayesian framework, this approach yields a \ac{map} estimation problem in which the regularization term describes the prior joint distribution postulated for the signals. In this respect, several studies employed~analytic and algorithmic tools, available in the literature of Bayesian estimation, 
to investigate the fundamental limits on the performance of \ac{dsn}s and design effective recovery schemes; see \cite{shiraki2016typical,zhu2017performance,hannak2017performance} and references therein.

When the source signals are spatially correlated, Bayesian estimation suggests to have a mutually coupled regularization term, since it can exploit the correlation among the signals. In general, the optimal performance is achieved when \ac{rls} uses the true prior distribution for regularization. The corresponding recovery algorithm is however not necessarily feasible to implement, e.g. in \ac{dcs}. For these cases, adopting a mismatched prior can lead to a computationally feasible \ac{rls} algorithm at the expense of slight degradation in the performance. An example of such mismatched~regularization is $\ell_{2,1}$-norm \ac{rls} for joint sparse recovery. In this algorithm, the regularization term is set to $\ell_{2,1}$-norm which for $[\bv_1, \ldots, \bv_J]$, with $\bv_j \in \setR^{N\times 1}$, is given by\cite{deng2013group}
\begin{align}
\norm{[\bv_1, \ldots, \bv_J]}_{2,1} = \sum_{n=1}^N \sqrt{ \sum_{j=1}^J \abs{v_{jn}}^2 }. \label{eq:l21}
\end{align}
For joint sparse recovery, $\ell_{2,1}$-norm is known to outperform classical regularization terms in compressive sensing, such~as $\ell_1$-norm. This observation follows the fact that $\ell_{2,1}$-norm, in contrast to classical regularization, does not assume independency among the jointly sparse signals. Although $\ell_{2,1}$-norm extracts the correlation among signals, the efficiency of the performance of this regularization is still questionable.~In~fact, the suggestion of such a regularization term is mainly based on heuristics and does not guarantee its superiority to other regularization terms with same computational complexity. 

%
\subsection*{Contributions}
In this paper, we characterize the performance of a generic form of \ac{map} estimation in \ac{dsn}s. Our investigations extends the scope of asymptotic decoupling principle, studied in \cite{rangan2012asymptotic,bereyhi2016rsb} for sensing networks with a single source, to settings with multiple terminals. Using this characterization, we propose the ``multi-dimensional soft thresholding'' algorithm for recovery of jointly sparse signals by imposing a more realistic postulation on the prior distribution. The proposed algorithm is posed as a linear programming and is shown to outperform $\ell_{2,1}$-norm regularization in terms of estimation error. This result indicates that given \ac{rls} algorithms with feasible computational complexity, $\ell_{2,1}$-norm is not in general the most efficient choice of regularization for joint sparse recovery.

\subsection*{Notations}
We represent scalars, vectors and matrices with non-bold, bold lower case and bold upper case letters, respectively. A $K \times K$ identity matrix is shown by $\mI_K$, and the $K \times K$ matrix and $K\times 1$ vector with all entries equal to one is denoted by $\mone_K$ and $\mone_{K\times 1}$, respectively. $\mH^{\trp}$ indicates the transpose of the matrix $\mH$. The set of real and integer numbers are denoted by $\setR$ and $\setZ$, and their corresponding non-negative subsets by superscript $+$. We denoted the Euclidean and $\ell_1$-norm with $\norm{\cdot}$ and $\norm{\cdot}_1$, respectively. For a given random variable $x$, either the probability mass or density function is represented with $\rmp(x)$. Moreover, $\E \cdot$ identifies the expectation operator. We use the shortened notation $[N]$ to represent $\set{1, \ldots , N}$.
\section{Problem Formulation}
\label{sec:sys}
We consider a general \ac{dsn} in which $J$ correlated source vectors, $\bx_j \in \setX^{N\times 1}$ for $j\in [J]$ and $\setX\subset \setR$,~are~measured linearly and individually as $\by_j=\mA_j\bx_j+\bz_j$ and received at a single data-fusion center. It is assumed that $\bx_j$, the sensing matrices $\mA_j \in \setR^{M_j \times N}$~and~measurement noises $\bz_j\in \setR^{N\times 1}$ satisfy the following constraints:

\begin{inparaenum}
\item[(a)] $\bx_1,\ldots,\bx_J$ are \ac{iid} such that the $n$-th sample of the source~terminals~for each $n\in[N]$ are spatially correlated and have~a~joint~probability distribution $\rmp_X(x_{n}^J)$ where $x_n^J\coloneqq (x_{1n}, \ldots, x_{Jn})$, i.e., 
\begin{align}
\rmp(\bx_1,\ldots,\bx_J) = \prod_{n=1}^N \rmp_X(x_n^J).
\end{align}

\item[(b)] $\mA_j \in \setR^{M_j \times N}$ for each $j\in[J]$ is randomly generated such that the Gram matrix $\mJ_j=\mA_j^{\trp} \mA_j$ has the decomposition $\mJ_j= \mU_j \mD_j \mU_j^{\trp}$ with $\mU_j$ being a Haar distributed matrix and $\mD_j$ denoting the diagonal matrix of eigenvalues. It is assumed that $\mA_j$ is independent of $\mA_k$ for any $j\neq k$, and the empirical distribution of eigenvalues, i.e., density of states, converges as $N \uparrow \infty$ to a deterministic~distribution $\rmp_j(\lambda)$. 

For the asymptotic distribution $\rmp_j(\lambda)$, the Stieltjes transform is given by $\rmG_j(s)= \E{(\lambda-s)^{-1}}$ for some complex $s$~with $\img{s} \geq 0$ where $\img{s}$ is the imaginary part of $s$. The $\rmR$-transform is moreover defined as 
\begin{align}
\rmR_j (\omega) = \rmG_j^{-1} (-\omega) - \omega^{-1}
\end{align}
such that $\lim_{\omega\downarrow 0} \rmR_j (\omega)\hspace*{-.7mm} =\hspace*{-.7mm} \E \lambda$ where $\rmG_j^{-1} (\cdot)$ denotes the inverse of the Stieltjes transform with respect to composition. This notation is further extended to matrix arguments: For the matrix $\mM_{N \times N}$ with the eigendecomposition $\mM=\mathbf{\Sigma}\mathbf{\Lambda} \mathbf{\Sigma}^{-1}$, $\mathrm{R}_j(\mM) \coloneqq \mathbf{\Sigma} \ \mathrm{diag}[\mathrm{R}_j(\lambda_1), \ldots, \mathrm{R}_j(\lambda_n)] \ \mathbf{\Sigma}^{-1}$. We use the $\rmR$-transform later to represent the main results.

%
\item[(c)] We consider a sequence of \ac{dsn}s with $N$ signal samples and $M_j$ measurements at each terminal. It is assumed that $M_j$, for $j\in[J]$, is a deterministic sequence of $N$ such that\vspace*{-1mm} 
\begin{align}
\rho_j \coloneqq \lim_{N \uparrow \infty} \frac{M_j}{N} < \infty.
\end{align}
We refer to $\rho_j$ as the $j$-th terminal compression ratio.

\item[(d)] $\bz_j \in \setR^{M_j\times 1}$ is an \ac{iid} zero-mean Gaussian random~ve- ctor with variance $\sigma_j^2$, i.e., $\bz \sim \man(\boldsymbol{0},\sigma_j^2 \mI_{M_j})$.

\end{inparaenum}
\subsection{Generic Joint \ac{map} Estimation}
\label{sec:Joint_MAP}
Although the measurements $\by_j$ are taken individually, the sources are reconstructed at a single fusion center. For recovery, a \ac{map} estimation algorithm is employed in which 
\begin{align}
\bhx^J = \argmin_{\bv^J} \sum_{j=1}^J \ &\frac{1}{2\lambda_j} \norm{\by_j-\mA_j\bv_j}^2 + u_\vv(\bv^J). \label{eq:MAP}
\end{align}
In \eqref{eq:MAP}, $\bhx^J \coloneqq (\bhx_1, \ldots, \bhx_J )$ denotes the recovered ensemble with $\bhx_j$ being the reconstruction of $\bx_j$. $u_\vv(\cdot) : \setR^{JN\times 1} \mapsto \setR^+$ describes a generic postulated prior and is referred to as the utility function. $\lambda_1, \ldots, \lambda_J$ are positive tuning factors which correspond to the assumed noise level in the original Bayesian inference problem and 
$\bv^J \coloneqq (\bv_1, \ldots, \bv_J)$ with minimization being taken over $\bv_j \in \setX^{N\times 1}$ for $j\in[J]$. As the source~samp- les are \ac{iid}, we consider a decoupling utility~function,~i.e., 
\begin{align}
u_\vv(\bv^J)=\sum_{n=1}^N u(v^J_{n})
\end{align}
for some $u(\cdot): \setR^{J\times 1} \mapsto \setR^+$ that takes into account the spatial correlation among the signal samples across the terminals.  
\subsection{Performance Measure}
The common metric to quantify the estimation performance is the \ac{mse} which determines the~distortion between the source vectors and their reconstructions averaged over all samples using the Euclidean distance. Nevertheless, the distortion metric can be defined for a generic measure. 
We therefore consider a general distortion function $\sfd(\cdot;\cdot): \setR^J \times \setR^J \mapsto \setR$ and %
define the distortion between the source ensemble $\bx^J$ and its reconstruction $\bhx^J$ as
\begin{align}
\sfd(\bhx^J; \bx^J)=\sum_{n=1}^{N} \sfd(\hx_{n}^J; x_{n}^J).
\end{align}
The average distortion is then given by
\begin{align}
\sfD_N = \frac{1}{N} \E{\sfd(\bhx^J ; \bx^J)}, \label{eq:dist}
\end{align}
and its ``asymptotic'', denoted by $\sfD$, is defined to be~the~limit when $N\uparrow \infty$. We intend to determine the asymptotic average distortion for the \ac{map} estimator presented in Section~\ref{sec:Joint_MAP}. As it is known from the literature, the direct approach~to~calculate $\sfD$ often fails as the optimization problem in \eqref{eq:MAP} does not have a closed-form solution for several choices of the utility function $u(\cdot)$ and source alphabet $\setX$. Moreover, the algorithmic~appr- oaches become computationally infeasible as $N$ increases. We thus invoke the replica method 
to determine $\sfD$.
\section{Main Results}
\label{sec:result}
In the large-system limit, the \ac{dsn} exhibits the decoupling property. This means that the network is statistically equivalent to $N$ identical scalar sensing networks. In the sequel, we state the decoupling property. The derivations are skipped here and postponed to Section~\ref{sec:large}. The validity of this result is based on the ``replica continuity'' conjecture which is later illustrated through the analyses in Section~\ref{sec:large}. In general, using the replica method, the asymptotic distortion $\sfD$ is derived as a solution of fixed-point equations referred to as the ``general replica ansatz''. When the setting exhibits  the so-called ``replica symmetry'' property\footnote{We discuss briefly the concept of replica continuity and replica symmetry in Section~\ref{sec:large}. More detailed discussions can be found in \cite{bereyhi2016statistical}.}, the general replica ansatz takes on a simple form. For simplicity, we assume in this section that the setting shows replica symmetry, and defer the presentation of the general replica ansatz to Section~\ref{sec:large}. From the literature \cite{bereyhi2016statistical,reeves2016replica}, it is well-known that \ac{map} estimators exhibit replica symmetry for a large class of utility functions. For specific cases, in which replica symmetry does not hold, the asymptotic distortion can be derived by following the systematic approach given in \cite{bereyhi2016statistical} to break replica symmetry.

\subsection{Decoupling Property}
\label{sec:decoupling}
To illustrate the decoupling property, we define the following single-letter ``decoupled sensing network'': The decoupled sensing network consists of scalar sources $\xx^J \coloneqq (\xx_1, \ldots, \xx_J)$ which are distributed with $\rmp_X(\xx^J)$ and measured for $j\in[J]$ as $\yy_j \hspace*{-.5mm} = \hspace*{-.5mm} \xx_j \hspace*{-.5mm} + \hspace*{-.5mm} \zz_j$ with independent~measurement noises $\zz_j\sim\man(0,\theta_j^2)$ whose variances 
are given by
\begin{align}
\theta_j^2 = \left[ \rmR_j(-\frac{\chi_j}{\lambda_j})\right]^{-2} \frac{\partial}{\partial\chi_j} \left[(\sigma_j^2 \chi_j-\lambda_j \sfp_j)\rmR_j(-\frac{\chi_j}{\lambda_j})\right] 
\end{align}
for some $\chi_1 , \ldots, \chi_J$ and $\sfp_1 , \ldots, \sfp_J$. At the fusion center, the sources are recovered via the single-letter \ac{map} estimator 
\begin{align}
\hxx^J = \argmin_{v^J} \sum_{j=1}^J \frac{1}{2\tau_j} (\yy_j-v_j)^2 + u(v^J). \label{eq:dec_est}
\end{align}
where $\tau_j$ is given by $\tau_j = {\lambda_j}/\rmR_j(-{\chi_j}/{\lambda_j})$. 


The decoupling property indicates that as $N$ increases, the pair $(x^J_n,\hx_{n}^J)$ for any index $n\in [N]$ converges in distribution to $(\xx^J, \hxx_n^J)$ when $\sfp_j= \E{ (\hxx_j-\xx_j)^2 }$ and $\chi_j$ satisfies
\begin{align}
\theta_j^2 \hspace*{.3mm} \chi_j = \tau_j \E{ \left(\hxx_j-\xx_j\right) \hspace*{.3mm} \zz_j} \label{eq:fix2}
\end{align}
for $j\in[J]$. Note that this property guarantees decoupling~on the marginal distribution $(x^J_n,\hx_{n}^J)$ and does not imply independency of recovered samples. In fact, $\hx^J_n$ and $\hx_{k}^J$ are in general correlated for $k\neq n$ \cite{rangan2012asymptotic,bereyhi2016rsb}.
\subsection{Asymptotic Average Distortion}
To calculate $\sfD$, we start from \eqref{eq:dist} and write 
\begin{align}
\sfD &= \lim_{N\uparrow \infty} \frac{1}{N} \E{\sfd(\bhx^J ; \bx^J)}= \lim_{N\uparrow \infty} \frac{1}{N} \sum_{n=1}^N \E{\sfd(\hx_n^J ; x_n^J)} \nonumber\\
&\stackrel{\dagger}{=} \lim_{N\uparrow \infty} \frac{1}{N} \sum_{n=1}^N \E{\sfd(\hxx^J ; \xx^J)}= \E{\sfd(\hxx^J ; \xx^J)}
\end{align}
where $\dagger$ comes from the decoupling property for large $N$.

\section{Distributed Compressive Sensing}
The most appealing application of the results is the problem of \ac{dcs}. Considering the generic form of the \ac{map}~estimator, the setup in Section~\ref{sec:sys} encloses a large class of \ac{dcs} settings with \ac{rls} recovery. In this section, we invoke Bayesian inference and propose the ``multi-dimensional soft thresholding''~al- gorithm. Using the asymptotic results, we show that this algorithm outperforms the well-known $\ell_{2,1}$-norm \ac{rls} recovery scheme while posing same computational complexity.
\subsection{Sparse Gaussian Priors}
\label{sec:prior}
We consider $J=2$ linearly-correlated jointly sparse sources with sparse Gaussian priors, i.e., $x_j$ for $j=1,2$ is written as
\begin{align}
x_j&=w_C s_C + w_j s_j
\end{align}
where $w_C\sim\man(0,\upsilon_C^2)$ and $w_j\sim\man(0,\upsilon_j^2)$, and $s_C$ and $s_j$ are Bernoulli distributed random variables with $\Pr\set{s_C=1}=1-\Pr\set{s_C=0}=\mu_C$ and $\Pr\set{s_j=1}=1-\Pr\set{s_j=0}=\mu_j$; moreover, $(w_C,w_1,w_2,s_C,s_1,s_2)$ are jointly independent. In this model, $w_C$ represents the part in the source signals that creates spatial correlation among samples; $s_C$ corresponds to the common support among the jointly sparse sources, and $w_j s_j$ denotes the part in each sparse source that is independent of the the signals at other terminals.
\subsection{Multi-dimensional Soft Thresholding Recovery}
The two-dimensional soft thresholding scheme recovers the correlated sparse sources by solving the \ac{map}~estimation~problem with the utility function
\begin{align}
u_\vv (\bv_1, \bv_2) = \norm{\bv_1}_1 + \norm{\bv_2}_1 + \psi \norm{\bv_1-\bv_2}_1 \label{utility_dif}
\end{align}
for some $\psi\geq 0$. 
The intuition behind such a utility function comes from the stochastic model of the jointly sparse sources. In fact, the model implies that $\bx_1-\bx_2$ also represents a sparse source which is linearly correlated to $\bx_1$ and $\bx_2$. This new vector can even be sparser that $\bx_1$ and $\bx_2$ when $\mu_C > \mu_1 , \mu_2$. We therefore invoke the sparsity of the difference term to take the spatial correlation of the source signals into consideration and add the term $\psi \norm{\bv_1-\bv_2}_1$ to the postulated prior. As the utility function is convex, the recovery algorithm is posed as a linear programming and efficiently solved. For more than two sources, the utility function is designed by a similar approach.



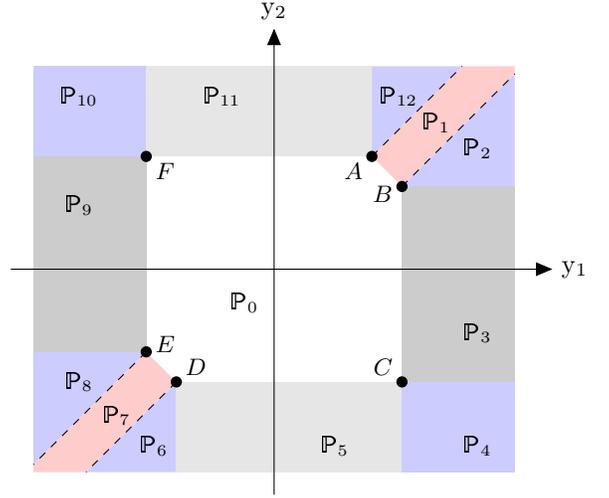
\begin{figure}
\hspace{10mm}
\begin{tikzpicture}
\begin{scope}

\fill[blue!20] (1.7,1.1) to (3.2,2.6) to (3.2,1.1) to (1.7,1.1);
\fill[blue!20] (1.3,1.5) to (1.3,2.7) to (2.5,2.7) to (1.3,1.5);

\fill[blue!20] (-1.7,-1.1) to (-3.2,-2.6) to (-3.2,-1.1) to (-1.7,-1.1);
\fill[blue!20] (-1.3,-1.5) to (-1.3,-2.7) to (-2.5,-2.7) to (-1.3,-1.5);

\fill[blue!20] (-1.7,1.5) to (-1.7,2.7) to (-3.2,2.7) to (-3.2,1.5) to (-1.7,1.5);
\fill[blue!20] (1.7,-1.5) to (1.7,-2.7) to (3.2,-2.7) to (3.2,-1.5) to (1.7,-1.5);

\fill[red!20] (-1.7,-1.1) to (-3.2,-2.6) to (-3.2,-2.7) to (-2.5,-2.7) to (-1.3,-1.5) to (-1.7,-1.1);
\fill[red!20] ( 1.7,1.1) to (3.2,2.6) to (3.2,2.7) to (2.5,2.7) to (1.3,1.5) to (1.7,1.1);

\fill[black!10] (-1.7,1.5) to (-1.7,2.7) to (1.3,2.7) to (1.3,1.5) to (-1.7,1.5);
\fill[black!10] (1.7,-1.5) to (1.7,-2.7) to (-1.3,-2.7) to (-1.3,-1.5) to (1.7,-1.5);

\fill[black!20] (1.7,1.1) to (1.7,-1.5) to (3.2,-1.5) to (3.2,1.1) to (1.7,1.1);
\fill[black!20] (-1.7,-1.1) to (-1.7,1.5) to (-3.2,1.5) to (-3.2,-1.1) to (-1.7,-1.1);
\end{scope}

\draw [->,>=triangle 45] (-3.5,0) -- (3.7,0) node [right] {$\yy_1$};
\draw [->,>=triangle 45] (0,-3) -- (0,3.2) node [above] {$\yy_2$};
\draw [dashed] (1.7,1.1) to (3.2,2.6);
%
\draw [dashed] (1.3,1.5) to (2.5,2.7);
\draw [dashed] (-1.7,-1.1) to (-3.2,-2.6);
%
\draw [dashed] (-1.3,-1.5) to (-2.5,-2.7);
%

\node[draw=none,shape=circle,fill, inner sep=1.5pt] (A) at (1.3,1.5) {}; 
\node at (1.3,1.3) [left] {\small{$A$}};

\node[draw=none,shape=circle,fill, inner sep=1.5pt] (B) at (1.7,1.1) {}; 
\node at (1.7,1) [left] {\small{$B$}};

\node[draw=none,shape=circle,fill, inner sep=1.5pt] (D) at (-1.3,-1.5) {}; 
\node at (-1.3,-1.3) [right] {\small{$D$}};

\node[draw=none,shape=circle,fill, inner sep=1.5pt] (E) at (-1.7,-1.1) {}; 
\node at (-1.7,-1) [right] {\small{$E$}};

\node[draw=none,shape=circle,fill, inner sep=1.5pt] (F) at (-1.7,1.5) {}; 
\node at (-1.7,1.3) [right] {\small{$F$}};

\node[draw=none,shape=circle,fill, inner sep=1.5pt] (C) at (1.7,-1.5) {}; 
\node at (1.7,-1.3) [left] {\small{$C$}};

\node at (2.15,1.7) [above] {\small{$\setP_1$}};
\node at (2.7,1.35) [above] {\small{$\setP_2$}};
\node at (2.7,-1.1) [above] {\small{$\setP_3$}};
\node at (2.7,-2.6) [above] {\small{$\setP_4$}};
\node at (.8,-2.6) [above] {\small{$\setP_5$}};
\node at (-1.6,-2.6) [above] {\small{$\setP_6$}};
\node at (-2.1,-2.2) [above] {\small{$\setP_7$}};
\node at (-2.6,-1.75) [above] {\small{$\setP_8$}};
\node at (-2.6,.6) [above] {\small{$\setP_{9}$}};
\node at (-2.6,2.05) [above] {\small{$\setP_{10}$}};
\node at (-.7,2.05) [above] {\small{$\setP_{11}$}};
\node at (1.65,2.05) [above] {\small{$\setP_{12}$}};

\node at (-.4,-.7) [above] {\small{$\setP_0$}};
\end{tikzpicture}
\caption{The thresholding regions for two-dimensional soft thresholding. In the blue regions, $\hxx_1$ and $\hxx_2$ are estimated as shifted versions of their corresponding observations while in the gray regions either $\hxx_1$ or $\hxx_2$ is estimated zero. When $(\yy_1,\yy_2)$ lies in the white region, both $\hxx_1$ and $\hxx_2$ are clipped to zero, and in the red region only $\hxx_1-\hxx_2$ is set to be zero meaning that $\hxx_1=\hxx_2$.\vspace*{-4mm}}
\label{fig:1}
\end{figure}
We now invoke the results in Section~\ref{sec:result} to characterize the asymptotic performance of two-dimensional~soft~thresholding recovery. The decoupled network illustrated in Section~\ref{sec:decoupling} for this scheme is illustrated using Fig.~\ref{fig:1}. Here, we have
\begin{figure}[t]
\hspace*{-.9cm}  
\resizebox{1.13\linewidth}{!}{

\pstool[width=.35\linewidth]{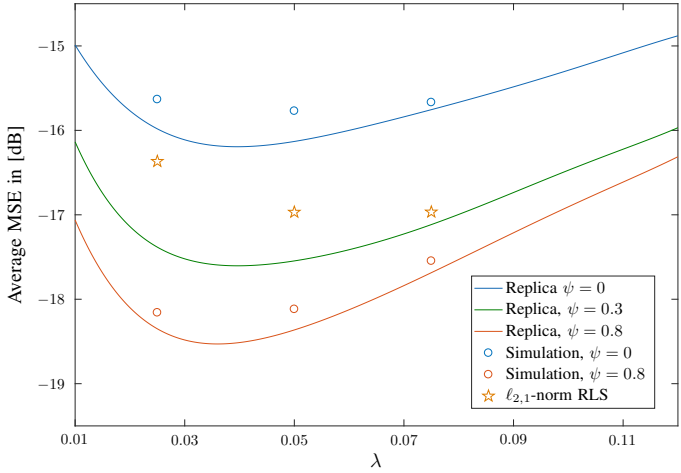}{

\psfrag{MSE}[c][t][0.26]{Average MSE in [dB]}
\psfrag{lam}[c][t][0.26]{$\lambda$}
\psfrag{PSI00REPBBBAAACCC}[l][l][0.22]{Replica $\psi=0$}
\psfrag{PSI03REPBBBAAACCC}[l][l][0.22]{Replica, $\psi=0.3$}
\psfrag{PSI08REPBBBAAACCC}[l][l][0.22]{Replica, $\psi=0.8$}
\psfrag{PSI00RESBBBAAACCC}[l][l][0.22]{Simulation, $\psi=0$}
\psfrag{PSI08RESBBBAAACCC}[l][l][0.22]{Simulation, $\psi=0.8$}
\psfrag{PLI20RESBBBAAACCC}[l][l][0.22]{$\ell_{2,1}$-norm RLS}
%
%
\psfrag{-19}[r][c][0.2]{$-19$}
\psfrag{-18}[r][c][0.2]{$-18$}
\psfrag{-17}[r][c][0.2]{$-17$}
\psfrag{-16}[r][c][0.2]{$-16$}
\psfrag{-15}[r][c][0.2]{$-15$}
%
\psfrag{0.01}[c][b][0.2]{$0.01$}
\psfrag{0.02}[c][b][0.2]{$0.02$}
\psfrag{0.03}[c][b][0.2]{$0.03$}
\psfrag{0.04}[c][b][0.2]{$0.04$}
\psfrag{0.05}[c][b][0.2]{$0.05$}
\psfrag{0.06}[c][b][0.2]{$0.06$}
\psfrag{0.07}[c][b][0.2]{$0.07$}
\psfrag{0.08}[c][b][0.2]{$0.08$}
\psfrag{0.09}[c][b][0.2]{$0.09$}
\psfrag{0.1}[c][][0.2]{$0.1$}
\psfrag{0.11}[c][b][0.2]{$0.11$}
\psfrag{0.12}[c][b][0.2]{$0.12$}

}
}
\caption{The average MSE vs. the tuning factor $\lambda$ for $\rho_1=\rho_2=0.8$. As the figure depicts, the two-dimensional soft thresholding scheme outperforms $\ell_{2,1}$-norm regularization.\vspace*{-4mm}}
\label{fig:2}
\end{figure}
\begin{subequations}
\begin{align}
A &= -D = [(1-\psi) \tau_1 ,(1+\psi)\tau_2 ]^\trp \\
B &= -E = [(1+\psi) \tau_1 ,(1-\psi)\tau_2 ]^\trp \\
C &= -F = [(1+\psi) \tau_1 ,-(1+\psi)\tau_2 ]^\trp 
\end{align}
\end{subequations}
where $\tau_1$ and $\tau_2$ represent the tuning factors in the decoupled network. The lines passing through $A$, $B$, $D$ and $E$ have unit slopes. As it is shown, the $\yy_1\yy_2$ plane is partitioned into 13 subsets $\setP_k$, for $k=0,\ldots,12$, using the points $A$ to $F$ and the lines indicated in the figure; for example, the partition $\setP_2$ is the set of all $\byy=[\yy_1,\yy_2]^\trp$ restricted among the line segment which connects $A$ to $B$ and the lines passing through $A$ and $B$. We now define the labeling functions $L_1(\cdot)$ and $L_2(\cdot)$ which assign a label to each point in the $\yy_1\yy_2$ plane.
\begin{align}
\hspace*{-2mm}
L_1(\byy) \hspace*{-.7mm} = \hspace*{-.7mm}
\begin{cases}
D_1 & \byy\in\setP_{12} \\
H_1 & \byy\in\setP_1 \\
S_1 & \byy\in\setP_{2:4}\\
D_2 & \byy\in\setP_{6}\\
H_2 & \byy\in\setP_{7}\\
S_2 & \byy\in\setP_{8:10}\\
O & \text{otherwise}
\end{cases}
\hspace*{-1.1mm}, \hspace*{.7mm}L_2(\byy) \hspace*{-.7mm} = \hspace*{-.7mm}
\begin{cases}
D_1 & \byy\in\setP_2 \\
H_1 & \byy\in\setP_{1} \\
S_1 & \byy\in\setP_{10:12}\\
D_2 & \byy\in\setP_{8}\\
H_2 & \byy\in\setP_{7}\\
S_2 & \byy\in\setP_{4:6}\\
O & \text{otherwise}
\end{cases}
\end{align}
where we have used the abbreviation~$\setP_{k:m} \hspace*{-.7mm}=\hspace*{-.7mm} \cup_{i=k}^m \setP_i$.~Using the labeling functions, the decoupled \ac{map} estimator is represented as $[\hxx_1,\hxx_2]=[\rmg_1(\byy), \rmg_2(\byy)]$ where
\begin{align}
\rmg_j(\byy) =
\begin{cases}
\yy_j+(-1)^\ell  (1-\psi) \tau_j & L_j(\byy)= D_\ell \\
\yy_j+(-1)^\ell  (1+\psi) \tau_j & L_j(\byy)= S_\ell \\
\dfrac{\yy_1 \tau_2+\yy_2 \tau_1+(-1)^\ell  2 \tau_1 \tau_2}{\tau_1+\tau_2} & L_j(\byy)= H_\ell \\
0 & L_j(\byy)= O
\end{cases}.\label{eq:decop}
\end{align}
From \eqref{eq:decop}, it is observed that the decoupled network is a~com- bination of three overlapping soft thresholding operators which operate on the $\yy_1$ and $\yy_2$ axes as well as the $\yy_1=\yy_2$ line\footnote{This observation clarifies further the appellation.}. The first two operators correspond to the terms $\norm{\bv_1}_1$ and $\norm{\bv_2}_1$ in the utility function while the latter thresholding is due to $\norm{\bv_1-\bv_2}_1$. As a result, the reconstructed symbols for $\byy\in \setP_1, \setP_7$ fulfill $\hxx_1=\hxx_2$ which correspond to clipping the single-letter term $\abs{v_1-v_2}_1$ to zero. By setting $\psi=0$ the points $A$ and $B$ as well as $D$ and $E$ meet, and the decoupled estimator reduces to two orthogonal soft thresholding operators.



\begin{figure}[t]
\hspace*{-.9cm}  
\resizebox{1.13\linewidth}{!}{

\pstool[width=.35\linewidth]{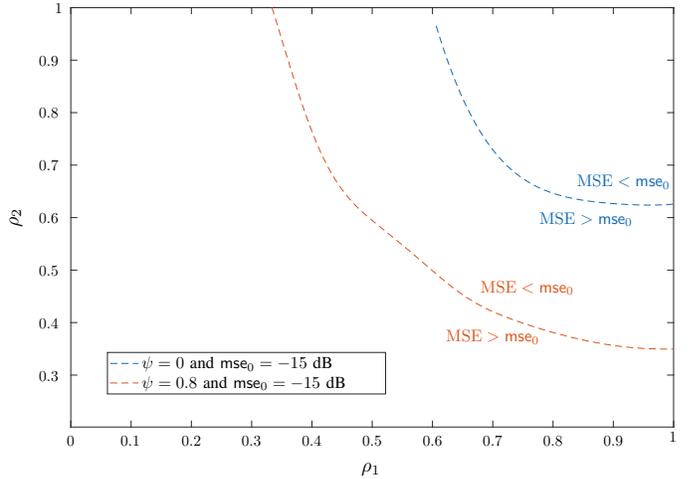}{

\psfrag{rho2}[c][b][0.26]{$\rho_2$}
\psfrag{rho1}[c][b][0.26]{$\rho_1$}
\psfrag{PSI00MSE15DBAAAABBBBCCCCD}[l][l][0.22]{$\psi=0$ and $\mse_0=-15$ dB}
\psfrag{PSI08MSE15DBAAAABBBBCCCCD}[l][l][0.22]{$\psi=0.8$ and $\mse_0=-15$ dB}

\psfrag{mse<mse1}[c][c][0.22]{\textcolor{myblue}{$\mathrm{MSE}<\mse_0$}}
\psfrag{mse>mse1}[c][c][0.22]{\textcolor{myblue}{$\mathrm{MSE}>\mse_0$}}

\psfrag{mse<mse2}[c][c][0.22]{\textcolor{myred}{$\mathrm{MSE}<\mse_0$}}
\psfrag{mse>mse2}[c][c][0.22]{\textcolor{myred}{$\mathrm{MSE}>\mse_0$}}

%
%
\psfrag{0.1x}[r][c][0.2]{$0.1$}
\psfrag{0.2x}[r][c][0.2]{$0.2$}
\psfrag{0.3x}[r][c][0.2]{$0.3$}
\psfrag{0.4x}[r][c][0.2]{$0.4$}
\psfrag{0.5x}[r][c][0.2]{$0.5$}
\psfrag{0.6x}[r][c][0.2]{$0.6$}
\psfrag{0.7x}[r][c][0.2]{$0.7$}
\psfrag{0.8x}[r][c][0.2]{$0.8$}
\psfrag{0.9x}[r][c][0.2]{$0.9$}
\psfrag{1x}[r][c][0.2]{$1$}
%
\psfrag{0}[c][b][0.2]{$0$}
\psfrag{0.1}[c][b][0.2]{$0.1$}
\psfrag{0.2}[c][b][0.2]{$0.2$}
\psfrag{0.3}[c][b][0.2]{$0.3$}
\psfrag{0.4}[c][b][0.2]{$0.4$}
\psfrag{0.5}[c][b][0.2]{$0.5$}
\psfrag{0.6}[c][b][0.2]{$0.6$}
\psfrag{0.7}[c][b][0.2]{$0.7$}
\psfrag{0.8}[c][b][0.2]{$0.8$}
\psfrag{0.9}[c][b][0.2]{$0.9$}
\psfrag{1}[c][][0.2]{$1$}

}
}
\caption{The rate-distortion region for two-dimensional soft thresholding with $\psi=0$ and $\psi=0.8$ when $\lambda=0.04$. The points above each diagram represent the pairs for which the \ac{mse} is~less~than~$\mse_0=$ $-15$ dB. By setting $\psi = 0.8$, the region significantly expands.\vspace*{-4mm}}
\label{fig:3}
\end{figure}
\subsection{Numerical Investigations}
For numerical investigations, we consider the case in which $\mA_1$\hspace*{-.8mm} and $\mA_2$\hspace*{-.4mm} have \ac{iid} zero-mean Gaussian entries~with~variance $1/M_j$. The $\rmR$-transform therefore reads $\rmR_j(\omega)={\rho_j}/({\rho_j- \omega})$.
We moreover set the distortion function to
\begin{align}
\sfd (\hx_1,\hx_2;x_1,x_2) = \frac 1 2 \norm{\hx_1-x_1}^2+ \frac 1 2 \norm{\hx_2-x_2}^2. \label{eq:dist}
\end{align}
As a benchmark, we further consider the $\ell_{2,1}$-norm~\ac{rls}~recovery scheme which is derived from the generic \ac{map} estimator in \eqref{eq:MAP} by setting $\ell_{2,1}$-norm in \eqref{eq:l21} as the utility function.

The plots in Fig.~\ref{fig:2} and Fig.~\ref{fig:3} are given for the prior model in Section~\ref{sec:prior} assuming that $\upsilon_C^2=\upsilon_j^2=0.5$, $\mu_C=0.3$ and $\mu_j=0.1$ for $j=1,2$. The noise variances are moreover set to $\sigma_j^2=0.01$ which means that the \ac{snr} for both terminals is $13$ dB. Due to the symmetry at the source terminals, we further set $\lambda_1=\lambda_2=\lambda$ in the simulations.

Fig.~\ref{fig:2} shows the average \ac{mse} of the network as a function of $\lambda$ for $\rho_1=\rho_2=0.8$ considering various choices of~$\psi$. From the figure, it is seen that the average \ac{mse} takes smaller values for $\psi\neq0$ which agrees with the intuition that for high spatial correlation among the source terminals the coupled prior in the proposed algorithm enhances the recovery performance. To validate the results given via the replica method, we have further sketched the average \ac{mse} calculated via numerical simulations for $N=100$ at some few points. For the sake of comparison, the simulation results for the $\ell_{2,1}$-norm \ac{rls} have been moreover indicated in the figure. One can observe that the proposed scheme outperforms $\ell_{2,1}$-norm recovery even for sub-optimal choices of $\psi$. Using the main results, one can further optimize the choice of $\psi$ and $\lambda_j$.

Fig.~\ref{fig:3} illustrates the rate-distortion region for $\lambda=0.04$ and the threshold distortion $\mse_0=-15$~dB~considering~$\psi=0$ and $\psi= 0.8$. For any rate pair $(\rho_1,\rho_2)$ above the diagram, the average \ac{mse} achieved by the corresponding recovery scheme is less than $\mse_0$. As the figure depicts, the region~expands~for $\psi= 0.8$ significantly following the fact that the proposed~alg- orithm takes the source correlation into consideration. 


\section{Sketch of the Derivations}
\label{sec:large}
In order to derive the decoupling property, we first modify the definition of the average distortion. Let $\setW_N\subset [N]$. Define the input-output distortion over the index set $\setW_N$ as
\begin{align}
\sfd^{\setW_N}(\bhx^J; \bx^J)=\sum_{n\in\setW_N} \sfd(\hx_{n}^J; x_{n}^J)
\end{align}
and set the asymptotic average distortion over the large limit of $\setW_N$ to $\sfD^\setW \hspace*{-.7mm} = \hspace*{-.7mm} \lim_{N\uparrow\infty}\mathbb{E}\{\sfd^{\setW_N}(\bhx^J; \bx^J)\}/\abs{\setW_N}$. We now define the function $\mae(\cdot|\cdot)$ which for given realizations of the quenched random set $\sfQ=\set{\mA^J,\by^J}$ reads
\begin{align}
\mae(\bv^J|\sfQ)= \sum_{j=1}^J \frac{1}{2\lambda_j} \norm{\by_j-\mA_j\bv_j}^2 + u(\bv^J).
\end{align}
By standard large deviations arguments, one can write 
\begin{align}
\sfD =\lim_{N\uparrow\infty} \lim_{\beta\uparrow\infty} \lim_{h\downarrow 0} \frac{\partial}{\partial h} \maf(\beta,h)
\end{align}
where $\maf(\beta,h)=-\mathbb{E} \log \maz(\beta,h)/N\beta $ with
\begin{align}
\maz(\beta,h) = { \sum_{\bv^J} \e^{-\beta \left[ \mae(\bv^J|\sfQ)+h \tfrac{N}{\abs{\setW_N}} \sfd^{\setW_N}(\bv^J;\bx^J)\right]}}. \label{eq:maf}
\end{align}
To bypass the hard task of integrating a logarithmic function, we utilize the Riesz equality and replace $\mathbb{E} \log \maz(\beta,h)$ with the term $\lim_{m\downarrow 0} \log \E{\maz^{m}(\beta,h)}/m$ in $\maf(\beta,h)$. The new expression represents $\maf(\beta,h)$ in terms of the moment function of $\maz(\beta,h)$. It is however not trivial to determine the moment function for real argument $m$. We thus invoke the ``replica continuity'' conjecture which assumes that the moment function analytically continues from $\setZ$ to $\setR$. This means that the moment function finds a same analytic form for both integer and real choices of $m$. 
In this case, by following the classic approach in \cite{bereyhi2016statistical}, after some lines of derivation we have
\begin{align}
\sfD^\setW = \lim_{m\downarrow 0} \lim_{\beta \uparrow \infty} \E{\sum_{\bvv^J} \sfd(\bvv^J;\bxx^J) \rmq_\beta(\bvv^J|\bxx^J) } \label{eq:general_ans}
\end{align}
where $\bvv^J$ and $\bxx^J$ are the ensembles of $\bvv_j=[v_{j1}, \ldots , v_{jm}]^\trp$ and $\bxx_j=x_j\mone_{m\times 1}$ for $j\in[J]$, respectively. The conditional distribution $\rmq_\beta(\bvv|\bxx)$ is moreover given by
\begin{align}
\rmq_\beta(\bvv|\bxx) = \frac{\e^{ -\beta\sum\limits_{j=1}^J (\bxx_j - \bvv_j)^\trp \mR_j(\bxx_j  - \bvv_j) + u(\bvv^J) }}{\sum_{\bvv^J} \e^{ -\beta\sum\limits_{j=1}^J (\bxx_j - \bvv_j)^\trp \mR_j (\bxx_j  - \bvv_j) + u(\bvv^J) }}
\end{align}
where $\mR_j \hspace*{-.5mm} \coloneqq \hspace*{-.5mm} \mT_j \rmR_{j} (-2\beta \mT_j \mQ_j)$ with $\mT_j \hspace*{-.5mm}=\hspace*{-.5mm} \frac{1}{2\lambda_j} (\mI_m - \frac{\beta\sigma_j^2}{\lambda_j} \mone_m)$ and the $m\times m$ matrices $\mQ_j$ for $j\in[J]$ satisfying
\begin{align}
\mQ_j &= \E{\sum_{\bvv} {(\bxx_j \hspace*{-.7mm} - \hspace*{-.7mm} \bvv_j) (\bxx_j \hspace*{-.7mm} - \hspace*{-.7mm} \bvv_j)^\trp} \rmq_\beta (\bvv|\bxx) }. \label{eq:fix_general}
\end{align}
The expression in \eqref{eq:general_ans} determines the general replica ansatz in terms of the correlation matrices $\mQ_j$. As $\rmq_\beta(\bvv|\bxx)$ is a function of $\mQ_j$, \eqref{eq:fix_general} gives a set of fixed-point equations whose solution determines the exact expression for the distortion.

Determining the explicit solution to \eqref{eq:fix_general} is not analytically possible, due to the replica continuity conjecture. We therefore invoke the approach in the literature of statistical mechanics and assume that the solution lies in a class of parameterized matrices. The most primary class is the replica symmetric set of matrices for which we have ${\mQ_j}= {\chi_j} \mI_m /{\beta} + \sfp_j \mone_m$.
For a large class of problems such a symmetry holds, and thus, the solution correctly determines the distortion. There are however some settings for which this assumption is not valid. For these cases, the replica symmetric structure should be generalized. 
Derivations for more general solutions can be followed in \cite{bereyhi2016statistical}.

Substituting the replica symmetric structure into the general ansatz, $\sfD^\setW$ is derived. As \eqref{eq:general_ans} does not depend on the index subset $\setW_N$, one can show that any joint moment of $(x_n^J,\hx_n^J)$ is equivalent to the corresponding joint moment of $(\xx^J,\hxx^J)$ following similar approach as in \cite{bereyhi2016rsb}. Invoking the moments method, the decoupling property is finally concluded.


\section{Conclusion}
This paper applied the replica method~to~study~the~performance of \ac{map} estimation in a distributed sensing framework. Our investigations extended the \ac{map} decoupling principle to these distributed setups. We further proposed a new algorithm for recovery of jointly~sparse sources and compared its performance with the traditional \ac{rls} approach in the literature. Numerical results depicted the superior performance of our algorithm compared with the $\ell_{2,1}$-norm \ac{rls} recovery scheme.

\bibliography{ref}
\bibliographystyle{IEEEtran}

\begin{acronym}
\acro{dsn}[DSN]{Distributed Sensing Network}
\acro{mmv}[MMV]{Multiple Measurement Vector}
\acro{dcs}[DCS]{Distributed Compressive Sensing}
\acro{mse}[MSE]{Mean Squared Error}
\acro{mimo}[MIMO]{Multiple-Input Multiple-Output}
\acro{csi}[CSI]{Channel State Information}
\acro{awgn}[AWGN]{Additive White Gaussian Noise}
\acro{iid}[i.i.d.]{independent and identically distributed}
\acro{ut}[UT]{User Terminal}
\acro{bs}[BS]{Base Station}
\acro{tas}[TAS]{Transmit Antenna Selection}
\acro{rls}[RLS]{Regularized Least Squares}
\acro{rhs}[r.h.s.]{right hand side}
\acro{lhs}[l.h.s.]{left hand side}
\acro{wrt}[w.r.t.]{with respect to}
\acro{rs}[RS]{Replica Symmetry}
\acro{rsb}[RSB]{Replica Symmetry Breaking}
\acro{papr}[PAPR]{Peak-to-Average Power Ratio}
\acro{rzf}[RZF]{Regularized Zero Forcing}
\acro{snr}[SNR]{Signal-to-Noise Ratio}
\acro{rf}[RF]{Radio Frequency}
\acro{map}[MAP]{Maximum-A-Posterior}
\acro{pmf}[PMF]{Probability Mass Function}
\acro{pdf}[PDF]{Probability Density Function}
\acro{cdf}[CDF]{Cumulative Distribution Function}
\end{acronym}
\end{document}